# CONTINUOUS GLUCOSE MONITORING PREDICTION

**Waggoner, Trae**
MCS, CIDSE
Ira A. Fulton Schools of Engineering
Arizona State University
tlwaggon@asu.edu

**Jose, Julia Ann**
MCS, CIDSE
Ira A. Fulton Schools of Engineering
Arizona State University
jajose2@asu.edu

**Manikandan B., Sudarsan**
MCS, CIDSE
Ira A. Fulton Schools of Engineering
Arizona State University
smanika1@asu.edu

## ABSTRACT
Diabetes is one of the deadliest diseases in the world and affects nearly 10 percent of the global adult population. Fortunately, powerful new technologies allow for a consistent and reliable treatment plan for people with diabetes. One major development is a system called continuous blood glucose monitoring (CGM). In this review we look at three different continuous meal detection methods that were developed given CGM data from patients with diabetes. From this analysis an initial meal prediction algorithm was also developed utilizing these methods.

## I.  INTRODUCTION

It is important to detect meal-intake in type-1 diabetic patients. Meal intake can affect glucose level of the body and it can even lead to harmful scenarios such as hyperglycemia. [2]

Fortunately, with the advancement of technology, there are different sensors and devices that let you monitor glucose levels. However, many of these devices come with their own limitations and inaccuracies. Thus, algorithmic detection can be a useful tool. This project focuses on 3 such algorithms for online meal detection and includes a suggestion for an initial algorithm as well. We focus on using an Auto-regression based model, Kalman-Filter based approach, an LSTM-RNN based approach.

We start off by first syncing the given CGM time series (corresponding to a patient) with the bolus ground truth. We then move to the development, instantiation, and implementation of the 3 algorithms. We then report the train and test accuracies. After which we go on to provide execution time analysis and on to the development of an initial algorithm for prediction of meal from CGM.

## II.  PROJECT SETUP

**Input:** CGM series and Bolus Data file which has a recording of glucose level of a person in 5 minute increments over 6 months. It had approximately 55000 records total with some empty values. The empty values were dropped and the rest of the values were considered.

Figure 1 illustrates the given CGM series for 1 day. Since the given records are in intervals of 5 minutes, this would mean 288 such records correspond to 1 day.

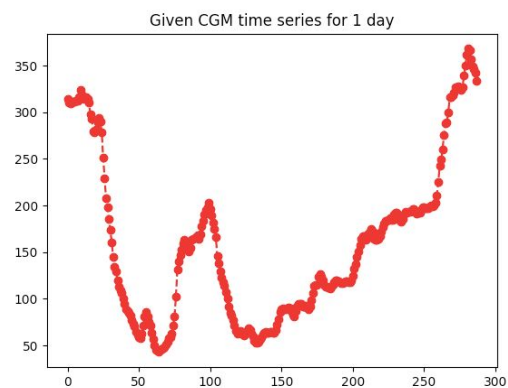

**Figure 1. CGM Series for 1 Day. The given graph shows CGM series data for the first 288 records or 1 day.**

**Task 1** of the project was to synchronize a given CGM series with ground truth. Figure 2 illustrates the CGM series synced with ground truth data.
As can be inferred from the graph, when there is a spike or slope in the CGM series there exists a spike in the bolus data as well, hence both the CGM as well as bolus series are synchronized.

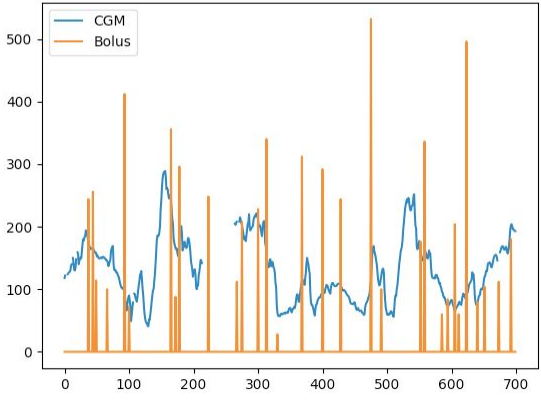

Figure 2. CGM vs. Ground Truth Graph

## III. IMPLEMENTATION DETAILS

We implemented three algorithms for meal detection: auto regression based model, kalman filter based model, and a Recurrent Neural Networks based approach. Each algorithm needed to be developed, instantiated, implemented, and then accuracy evaluated. This was done using training and testing data.

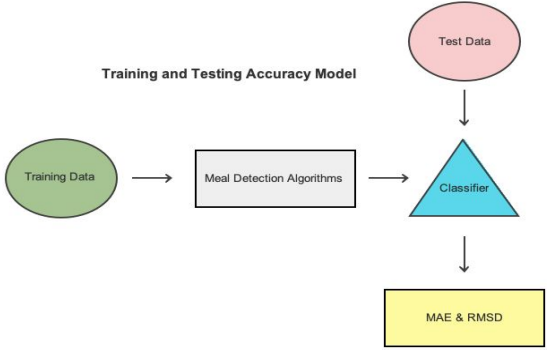

Figure 3. Training and Testing Accuracy Model

We used Mean Absolute Error and Root Mean Squared Error for error/loss calculation. The formulas are written below.

$$MAE = \frac{\sum_{i=1}^{n} |y_i - x_i|}{n}$$

MAE = mean absolute error
$y_i$ = prediction
$x_i$ = true value
n = total number of data points

$$RMSD = \sqrt{\frac{\sum_{i=1}^{N}(x_i - \widehat{x}_i)^2}{N}}$$

RMSD = root-mean-square deviation
i = variable i
N = number of non-missing data points
$x_i$ = actual observations time series
$\widehat{x}_i$ = estimated time series

### A. Auto Regression - SARIMA

Seasonal Autoregressive Integrated Moving Average (SARIMA) is a very common and useful technique for time series modelling and forecasting. The fact that SARIMA accounts for the seasonal variation in time series is what makes it so powerful. Our CGM data has seasonal components in it which explains the use of SARIMA for this project. SARIMA comes with trend and seasonality parameters that must be configured before you use it. The hyperparameters for trend elements like order of trend (p), difference order (d), the trend moving average order (q) and seasonal elements like Seasonal auto-regressive order (P), Seasonal differencing order (D), Seasonal moving average (Q) and seasonal period (m) need to be configured prior to using the model. To determine optimal values of these parameters, we used Autocorrelation and Partial Autocorrelation plots. Using these plots and the AIC (Akaike information criterion) scores, we were able to select optimal hyperparameters for our model. [3] The Autocorrelation and Partial Autocorrelation plots are shown in figure 4. Using these hyperparameters, we trained our model on the given CGM series.

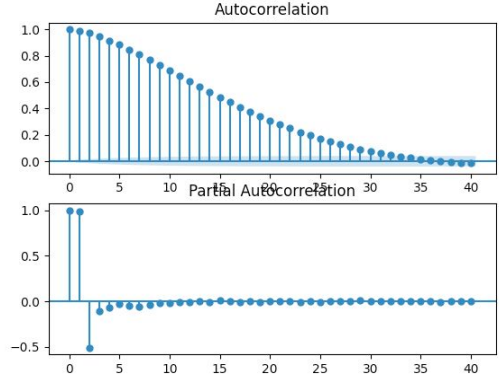

Figure 4. Autocorrelation and Partial Autocorrelation plots

### B. Kalman Filter

The Kalman Filter is an algorithm that takes in time series measurements that have varying levels of statistical noise and predicts accurately estimated values for unknown variables. This is precisely what we need to help people with diabetes, "an algorithm that can detect meals from real-time measured variables" [1]. The Kalman Filter is broken up into two important parts: update and predict. Using the CGM data we trained our Kalman Filter to effectively predict the patient's glucose levels.

In order to train the Kalman Filter, the given data is consecutively inputed in order to find the true value. We needed to use three main equations to get closer and closer to the true value: kalman gain, current estimate and error in estimate. First, we needed an initial estimate and an initial error in the estimate. From there we were able to perform our necessary calculations with each data point input. The equations for the 3 calculations are given below.

1. $Kalman\ Gain\ =\ \frac{E_E}{E_E + E_M}$

   $E_E$ = error in estimate
   $E_M$ = error in measurement

2. $Current\ Estimate\ =\ P_E + KG(m - P_E)$

   $P_E$ = previous estimate
   m = measurement
   KG = Kalman Gain

3. $Error\ in\ Estimate\ =\ [1 - KG](E_P)$

   $E_P$ = error in previous estimate

### C. Recurrent Neural Networks

We further implemented an RNN based model for meal detection. RNN usually suffers from what is known as "short memory" which is a result of the vanishing gradient problem [4] that neural networks usually encounter. For this reason, we implemented an LSTM (Long Short Term Memory) model. It overcomes the memory problem by using gates to control information flow.

For the LSTM model, we first split the dataset into train and test dataset. We then go on to create sub sequences of our input sets and train a Sequential LSTM model on it.

### D. Initial algorithm for meal prediction

Here, we developed an initial algorithm for meal prediction. This algorithm uses an LSTM model but slightly differently. In part C, for our implementation of the LSTM model, we trained the model on our given dataset which is the dynamically changing time series. In this part, we go on to add certain constant values that define our time series, while training the model. [5] Upon exploration of different feature engineering techniques, we stumbled upon certain transformations that made our given CGM series look more intuitive. An example would be the Fast Fourier Transformation [6] as shown in figure 5.

For a given CGM series, the FFT tends to have a high peak. Hence, the highest peak of FFT transformation of our series defines the CGM series. We exploited two more such transformations such as Continuous Wavelet Transform and Discrete Wavelet Transform. Figure 6 and 7 illustrates these transformations. We used several statistics such as the mean and variance of these transformations to define our CGM series.

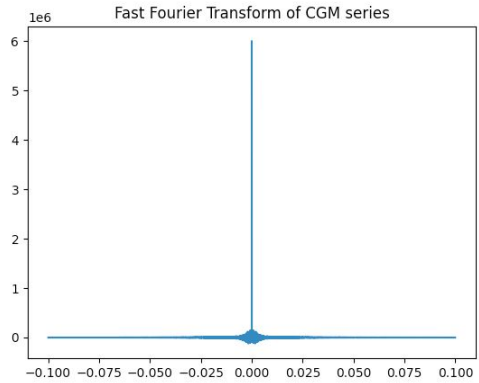

Figure 5. Fast Fourier Transform (FFT) of given CGM series

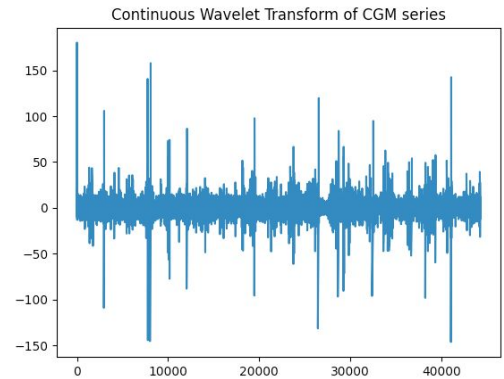

Figure 6. Continuous Wavelet Transform (CWT) of given CGM series

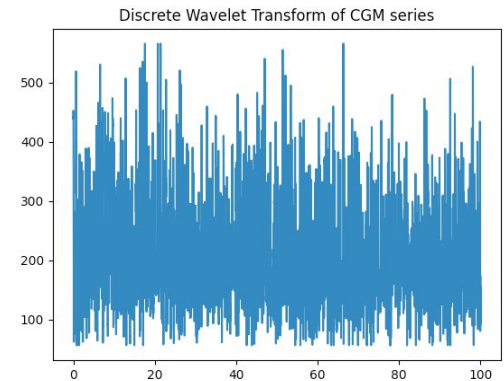

Figure 7. Discrete Wavelet Transform (DWT) of given CGM series

We let 7 such constants define our series:

1. FFT first highest peak
2. FFT second highest peak
3. FFT Mean
4. CWT Mean
5. CWT Variance
6. DWT Mean
7. DWT Variance

These 7 constants were passed to our CGM series data in LSTM. This was done by flattening the output of the LSTM layer and then concatenating these constant values to it and applying a dense layer to the model. Table 1 summarizes the train and test losses as reported by this algorithm. Figure 8 shows the train and test learning curves.

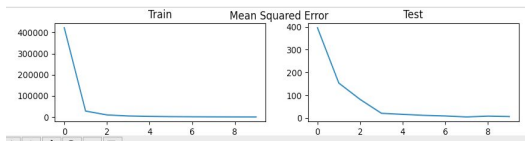

**Figure 8. Train and test MSE loss for modified LSTM model**

## IV. RESULTS AND FINDINGS

### A. Auto Regression - SARIMA

After using a 80-20 split of given dataset into train/test dataset, we trained the SARIMA model on train dataset and used the test dataset to test it. The train and test accuracy is summarized in table 1. We predicted the next 30 minutes, 1 hour as well as 2 hours of the series to get an intuition of meal intake. The predicted and actual time series values can be seen in figure 9.

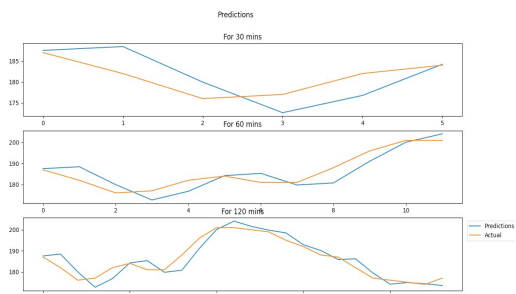

**Figure 9. SARIMA Meal Detection Graphs. The first graph illustrates the predicted versus actual glucose values of the given patient for the next 30 minutes. The second graph is for the next 1 hour. And the third graph shows the next 2 hours.**

We can see that the predicted values are very close to the actual values. Figure 10 shows the values of predicted and actual for the next 30 minutes.

```
   Predictions  Actual
0  187.534427   187.0
1  188.457023   182.0
2  179.951518   176.0
3  172.637117   177.0
4  176.763228   182.0
5  184.212135   184.0
```

**Figure 10. SARIMA Prediction results for next 30 minutes.**

### B. Kalman Filter

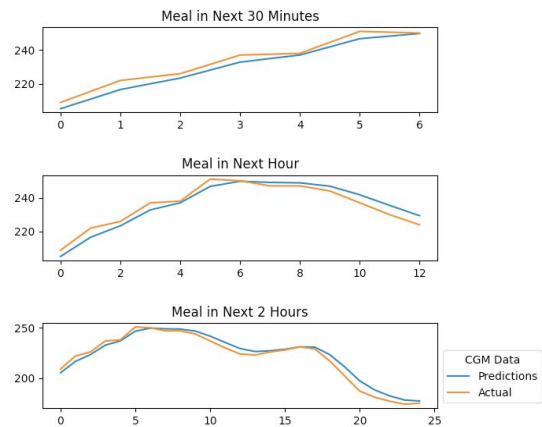

**Figure 11. Kalman Filter Meal Detection Graphs. The first graph illustrates the predicted versus actual glucose values of the given patient for the next 30 minutes. The second graph is for the next 1 hour. And the third graph shows the next 2 hours.**

We can see that the predicted values are, like the SARIMA algorithm, extremely consistent. Figure 12 shows the values of predicted and actual for the next 30 minutes.

```
   Predictions  Actual
0  205.312886   209.0
1  216.626089   222.0
2  223.419485   226.0
3  232.812705   237.0
4  237.018630   238.0
5  246.659592   251.0
6  249.724078   250.0
```

**Figure 12. Kalman Filter Prediction results for next 30 minutes.**

### C. Recurrent Neural Networks

The given dataset was used to train our LSTM model on. Our model consisted of a sequential model with a single layer of LSTM and a dropout layer. It was trained for 10 epochs and a validation test using 20% of the input training dataset. Table 1 summarizes the train and test accuracy of our model on the given datasets.

We predicted the next 30 minutes, 1 hour as well as 2 hours of the series to get an intuition of meal intake. Figure 13 illustrates the plots for mean squared error and mean absolute error for both train and test datasets.

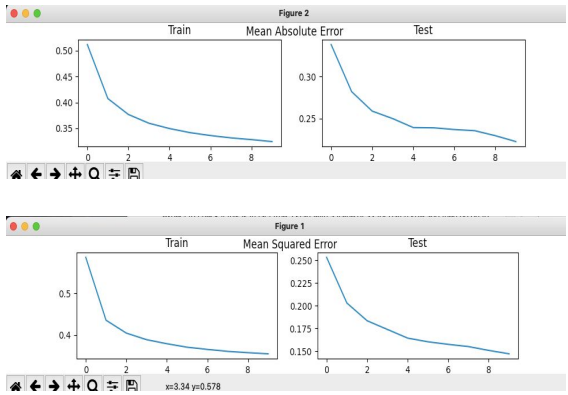

Figure 13. LSTM MAE and MSE plots

| Algorithm | Train | | Test | |
|---|---|---|---|---|
| | MAE | RMSE | MAE | RMSE |
| **SARIMA** | 3.87 | 6.90 | 3.45 | 4.16 |
| **KALMAN** | 3.21 | 4.36 | 3.08 | 3.79 |
| **RNN** | 0.32 | 0.56 | 0.22 | 0.47 |
| **Modified LSTM** | 11.0 | 20.15 | 5.43 | 7.05 |

Table 1. Train and Test losses: Mean Absolute Error, Roost Mean Squared Error

### D. Execution Time Analysis

The time taken for training for each approach is calculated in this analysis. SARIMA took less than 1 minute to train. Kalman filter took less than 2 seconds. RNN took less than 10 minutes. Finally, the modified-RNN took less than 10 minutes.

### V. LIMITATIONS

We were unable to get access to IMPACT lab and it's resources related to the implementation of kalman-filter based approach, however, we were able to implement it with the help of the paper [1].

### VI. CONCLUSION

In this project, we presented our strategies to predict the user's glucose level, indirectly indicating the amount and time the insulin is injected into the user. With the strategies; SARIMA, Kalman Filter, and Recurrent Neural Networks, we were able to achieve good performance with each model. Additionally, we implemented an initial algorithm for meal prediction, which involves an LSTM model along with certain constant values that define the time series, while training the model. From all the approaches we studied, we find the Recurrent Neural Networks to predict the time series values closest to the actual time series values.

### VII. ACKNOWLEDGEMENT

We would like to thank Prof Ayan Banerjee for providing us the opportunity to come together as a team and work on this interesting project.

### VIII. REFERENCES

1. Turksoy K, Samadi S, Feng J, Littlejohn E, Quinn L, Cinar A. Meal Detection in Patients With Type 1 Diabetes: A New Module for the Multivariable Adaptive Artificial Pancreas Control System. *IEEE J Biomed Health Inform*. 2016;20(1):47-54. doi:10.1109/JBHI.2015.2446413
2. Adam Felman. (May 7, 2019). What to know about Hyperglycemia. https://www.medicalnewstoday.com/articles/323699#:~:text=Hyperglycemia%20refers%20to%20high%20levels,cells%20for%20use%20as%20energy.
3. A Gentle Introduction to SARIMA for Time Series Forecasting in Python. https://machinelearningmastery.com/sarima-for-time-series-forecasting-in-python/

4. Vanishing gradient problem. https://en.wikipedia.org/wiki/Vanishing_gradient_problem
5. Adding static data (not changing over time) to sequence data in LSTM. https://stackoverflow.com/questions/53363986/adding-static-data-not-changing-over-time-to-sequence-data-in-lstm
6. Fast Fourier transform. https://en.wikipedia.org/wiki/Fast_Fourier_transform